\documentclass[aps,showpacs,floatfix,pre,twocolumn]{revtex4}

\usepackage{amsmath,graphicx,subfigure}

\renewcommand{\vec}[1]{\boldsymbol{\mathbf{#1}}}

\newcommand{\Ceps}{C_\varepsilon}
\newcommand{\Cinf}{C_{\varepsilon,\infty}}

\newcommand{\vep}{\varepsilon}
\newcommand{\RL}{R_L}
\newcommand{\Rl}{R_{\lambda}}

\newcommand{\beq}{\begin{equation}}
\newcommand{\eeq}{\end{equation}}

\begin{document}

\title{Analysis of the Taylor dissipation surrogate in forced isotropic 
turbulence}

 \author{W.~D. McComb}
 \affiliation{
 SUPA, School of Physics and Astronomy,
 University of Edinburgh, Edinburgh EH9 3JZ, UK}
 \author{A. Berera}
 \affiliation{
 SUPA, School of Physics and Astronomy,
 University of Edinburgh, Edinburgh EH9 3JZ, UK}
 \author{S.~R. Yoffe}
 \affiliation{
 SUPA, School of Physics and Astronomy,
 University of Edinburgh, Edinburgh EH9 3JZ, UK}

\begin{abstract}
From the energy balance in wavenumber space expressed by the Lin equation, 
we derive a new form for the local Karman-Howarth equation for forced 
isotropic turbulence in real space. This equation is then cast into a 
dimensionless form, from which a combined analytical and numerical study 
leads us to deduce a new model for the scale-independent nondimensional 
dissipation rate $\Ceps$, which takes the form $\Ceps = \Cinf +
C_L/R_L$, where the asymptotic value $\Cinf$ can be evaluated from the
third-order structure function. This is found to fit the numerical data with $\Cinf = 0.47 \pm 
0.01$ and $C_L= 18.5 \pm 1.3$. By considering $\Ceps - \Cinf$ on logarithmic 
scales, we show that $R_L^{-1}$ is indeed the correct Reynolds number behaviour. 
The model is compared to previous attempts in the literature, with 
encouraging agreement.  The effects of the scale-dependence of the inertial 
and viscous terms due to finite forcing are then considered and shown to 
compensate one another, such that the model equation is applicable for 
systems subject to finite forcing.
In addition, we also show that, contrary to the case of freely
decaying turbulence, the characteristic decline in $\Ceps$ with
increasing Reynolds number is due to the \emph{increase} in the
surrogate expression $U^3/L$; the dissipation rate being maintained constant as a
consequence of the fixed rate of forcing. A long-time non-turbulent stable 
state is found to exist for low Reynolds number numerical simulations which 
use negative damping as a means of energy injection.
\end{abstract}


\maketitle

\section{Introduction}

In recent years there has been much interest in the fundamentals of
turbulent dissipation. This interest has centred on the approximate
expression for the dissipation rate $\vep$ which was given by Taylor in
1935 \cite{Taylor35} as
\beq
\vep = \Ceps U^3/L,
\label{Taylor-diss}
\eeq
where $U$ is the root-mean-square velocity and $L$ is the integral
scale. Many workers in the field refer to equation (\ref{Taylor-diss})
as the \emph{Taylor dissipation surrogate}. However, others rearrange it
to define the coefficient $\Ceps$ as the nondimensional dissipation
rate, thus:
\beq
\Ceps = \frac{\vep}{U^3/L}.
\label{dim-diss}
\eeq
In 1953 Batchelor \cite{Batchelor71} presented evidence to
suggest that the coefficient $\Ceps$ tended to a constant value with
increasing Reynolds number. In 1984 Sreenivasan \cite{Sreenivasan84}
showed that in grid turbulence $\Ceps$ became constant for
Taylor-Reynolds numbers greater than about $50$. Later still, in 1998,
he presented a survey of investigations of both forced and decaying
turbulence \cite{Sreenivasan98}, using direct numerical simulation
(DNS), which established the now characteristic curve of $\Ceps$ plotted
against the Taylor-Reynolds number $\Rl$.

In his 1968 lecture notes \cite{Saffman68}, Saffman made two comments
about the expression that we have given here as equation
(\ref{Taylor-diss}). These were: ``This result is fundamental to an
understanding of turbulence and yet still lacks theoretical support''
and  ``the possibility that $A$ (\emph{i.e. our} $\Ceps$) depends weakly
on the Reynolds number can by no means be completely discounted''. More
than forty years on, the question implicit in his second comment has
been comprehensively answered by the survey papers of Sreenivasan
\cite{Sreenivasan84,Sreenivasan98}, along with a great deal of
subsequent work by others, some of which we have cited here. However,
while some theoretical work has indicated an inverse proportionality
between $\Ceps$ and Reynolds number, this has been limited to low (i.e.
non-turbulent) Reynolds numbers \cite{Sreenivasan84} or based on a
mean-field approximation \cite{Lohse94} or restricted to providing an
upper-bound \cite{Doering02}.  Hence his first comment is still valid
today; and this lack of theoretical support remains one of the main
impediments to the development of turbulence phenomenology and hence
turbulence theory.

As we have seen before, an approach based on the dimensionless
dissipation $\Ceps$, the ratio of the dissipation to the surrogate
expression $U^3/L$, can be a helpful way of looking at things
\cite{McComb10b}. In the present paper, we examine the behaviour of
$\Ceps$ with increasing Reynolds number by means of a simple model based
on the Karman-Howarth equation and supported by direct numerical
simulation (DNS). We find that this description captures the observed
dependence of $\Ceps$, thus providing a direct theoretical route from
the Navier-Stokes equation to dissipation rate scaling. We begin with a
description of our DNS, before presenting a theoretical analysis
followed by numerical  results.

\section{The numerical simulations}

We used a pseudospectral DNS, with full dealiasing performed by
truncation of the velocity field according to the two-thirds rule. Time 
advancement for the viscous term was performed exactly using an integrating 
factor, while the non-linear term used Heun's method (second-order 
predictor-corrector). Each
run was started from a Gaussian-distributed random field with a
specified energy spectrum (which behaves as $k^4$ for the low-$k$
modes), and was allowed to a reach steady-state before measurements were
made. A deterministic forcing scheme was employed, with the force
$\vec{f}$ given by
\begin{align}
 \vec{f}(\vec{k},t) &=
      (\varepsilon_W/2 E_f) \vec{u}(\vec{k},t) \quad
\text{for} \quad  0 < \lvert\vec{k}\rvert < k_f ; \nonumber \\
  &= 0   \quad \textrm{otherwise},
\label{forcing}
\end{align}
where $\vec{u}(\vec{k},t)$ is the instantaneous velocity field (in
wavenumber space). The highest forced wavenumber, $k_f$, was chosen to
be $k_f = 2.5$. As $E_f$ was the total energy contained in the forcing
band, this ensured that the energy injection rate was $\varepsilon_W =
\textrm{constant}$. It is worth noting that any method of energy
injection  employed in the numerical simulation of isotropic turbulence
is not experimentally realisable. The present method  of negative
damping has also been used in other investigations 
\cite{Jimenez93,Yamazaki02,Kaneda03,Kaneda06}, albeit not necessarily
such  that $\vep_W$ is maintained constant (although note the
theoretical analysis of this type of forcing by Doering and Petrov
\cite{Doering05}), and we stress that at no point do we  rely on the fact
that the force is correlated with the velocity.

For each Reynolds number studied, we used the same initial spectrum and
input rate $\varepsilon_W$. The only initial condition changed was the
value assigned to the (kinematic) viscosity. Once the initial transient
had passed the velocity field was sampled every half a large-eddy
turnover time, $\tau = L/U$. The ensemble populated with these sampled
realisations was used, in conjunction with the usual shell averaging, to
calculate statistics. Simulations were run using lattices of size
$64^3,\ 128^3,\ 256^3,\ 512^3$ and $1024^3$, with corresponding Reynolds
numbers ranging from $R_\lambda = 8.40$ up to $335.2$. The smallest
resolved wavenumber was $k_\text{min} = 2\pi/L_\text{box} = 1$ in all
simulations, while the maximum wavenumber always satisfied $k_\text{max}\eta
 > 1.0$, where $\eta$ is the Kolmogorov dissipation lengthscale.
The integral scale, $L$, was found to lie between $0.34 L_{\text{box}}$
and $0.18 L_{\text{box}}$. Details of the individual runs are summarised in 
table \ref{tbl:simulations}.

\begin{table}[tb!]
 \begin{center}
  \begin{tabular}{l|ll|lll|ll}
  $R_\lambda$ & $\nu_0$ & $N$ & $\varepsilon$ & $\sigma$ & $U$ & 
  $L/L_\text{box}$ & $k_\text{max}\eta$ \\
  \hline
  8.40  & 0.09    & 64   & 0.085 & 0.011 & 0.435 & 0.34 & 6.09 \\
  9.91  & 0.07    & 64   & 0.081 & 0.014 & 0.440 & 0.32 & 5.10 \\
  13.9  & 0.05    & 64   & 0.086 & 0.014 & 0.485 & 0.31 & 3.91 \\
  24.7  & 0.02    & 64   & 0.092 & 0.011 & 0.523 & 0.24 & 1.93 \\
  41.8  & 0.01    & 512  & 0.097 & 0.010 & 0.581 & 0.22 & 9.57 \\
  42.5  & 0.01    & 128  & 0.094 & 0.015 & 0.581 & 0.23 & 2.34 \\
  44.0  & 0.009   & 128  & 0.096 & 0.009 & 0.587 & 0.22 & 2.15 \\
  48.0  & 0.008   & 128  & 0.096 & 0.013 & 0.586 & 0.22 & 1.96 \\
  49.6  & 0.007   & 128  & 0.098 & 0.011 & 0.579 & 0.20 & 1.77 \\
  60.8  & 0.005   & 512  & 0.098 & 0.009 & 0.589 & 0.20 & 5.68 \\
  64.2  & 0.005   & 128  & 0.099 & 0.011 & 0.607 & 0.21 & 1.37 \\
  89.4  & 0.0025  & 512  & 0.101 & 0.006 & 0.605 & 0.19 & 3.35 \\
  101.3 & 0.002   & 256  & 0.099 & 0.009 & 0.607 & 0.19 & 1.41 \\
  113.3 & 0.0018  & 256  & 0.100 & 0.008 & 0.626 & 0.20 & 1.31 \\
  153.4 & 0.001   & 512  & 0.098 & 0.011 & 0.626 & 0.20 & 1.70 \\
  176.9 & 0.00072 & 512  & 0.102 & 0.009 & 0.626 & 0.19 & 1.31 \\
  203.7 & 0.0005  & 512  & 0.099 & 0.008 & 0.608 & 0.18 & 1.01 \\
  276.2 & 0.0003  & 1024 & 0.100 & 0.009 & 0.626 & 0.18 & 1.38 \\
  335.2 & 0.0002  & 1024 & 0.102 & 0.008 & 0.626 & 0.18 & 1.01
  \end{tabular}
 \end{center}
 \caption{A summary of the main parameters for our numerical simulations. 
 The values quoted for the dissipation rate $\varepsilon$ and its standard 
 deviation $\sigma$, the total energy $E$ and the velocity-derivative 
 skewness $S$, are ensemble- and shell-averaged mean values.}
 \label{tbl:simulations}
\end{table}

In addition, we note that all data fitting has been performed using an 
implementation of the nonlinear-least-squares Marquardt-Levenberg algorithm, 
with the error quoted being one standard error.

\begin{figure}
 \begin{center}
  \includegraphics[width=0.45\textwidth]{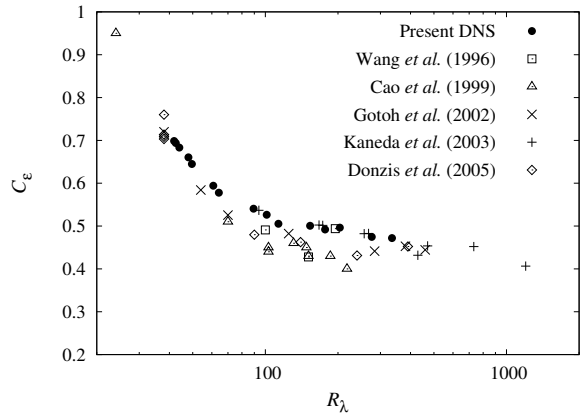}
 \end{center}
 \caption{ Variation of the dimensionless dissipation
coefficient with Taylor-Reynolds number. Other investigations of forced
turbulence are presented for comparison.}
 \label{fig:Ceps_Rl}
\end{figure}

Our simulations have been well validated by means of extensive and detailed
comparison with the results of other investigations. These include the 
Taylor-Green vortex \cite{Taylor37,Brachet83}; measurements of the isotropy, 
Kolmogorov constant and velocity-derivative skewness; advection of a passive 
scalar; and a direct comparison with the freely-available pseudospectral 
code \emph{hit3d} \footnote{S. Chumakov, N. Vladimirova, and M. Stepanov.  
Available from: \texttt{http://code.google.com/p/hit3d/}.}.
These will be
presented in another paper, but it can be seen from Fig.
\ref{fig:Ceps_Rl} that our results reproduce the characteristic
behaviour for the plot of $\Ceps$ against $\Rl$, and agree closely with
other representative results in the literature
\cite{Wang96,Cao99,Gotoh02,Kaneda03,Donzis05}. We note that the data 
presented for comparison was obtained using negative-damping (with variable 
$\vep_W$) \cite{Kaneda03}, stochastic noise \cite{Gotoh02,Donzis05}, or 
maintaining a $k^{-5/3}$ energy spectrum within the forced shells 
\cite{Wang96,Cao99}. These methods for energy injection have been discussed 
in \cite{Bos07}.

\section{A Dimensionless Karman-Howarth equation for forced turbulence}

The use of stirring forces with the energy equation in spectral space
(i.e. with the Lin equation) is well established,
\begin{equation}
 \frac{\partial E(k)}{\partial t} = T(k) - 2\nu_0 k^2 + W(k) \ ,
\end{equation}
where $\nu_0$ is the kinematic viscosity, $E(k)$ and $T(k)$ are the energy 
and transfer spectra, respectively, and $W(k) = 4\pi k^2 \langle 
\vec{u}(-\vec{k})\cdot\vec{f}(\vec{k}) \rangle$ is the work spectrum of the 
stirring force, $\vec{f}(\vec{k})$.
(See, for example,
\cite{McComb90a}.)  But this is not the case with the Karman-Howarth
equation (KHE), which is its real-space equivalent. Accordingly, we obtain 
the
equivalent KHE by Fourier transformation of the Lin equation (with
forcing) as
\begin{align}
 \label{gen-khe}
 -\frac{3}{2}\frac{\partial U^2}{\partial t} +\frac{3}{4}\frac{\partial 
 S_2(r)}{\partial t} &= -\frac{1}{4r^4} \frac{\partial}{\partial r} \Big( 
 r^4 S_3(r) \Big) \\
 &\qquad+ \frac{3\nu_0}{2r^4} \frac{\partial}{\partial
r}\left( r^4 \frac{\partial S_2(r)}{\partial r} \right)
 - I(r) \ , \nonumber
\end{align}
where the longitudinal structure functions are defined as
\begin{equation}
 S_n(r) = \left\langle \big([\vec{u}(\vec{x}+\vec{r}) - 
 \vec{u}(\vec{x})]\cdot\vec{\hat{r}}\big)^n \right\rangle \ .
\end{equation}
The input $I(r)$ is given in terms of $W(k)$, the work spectrum of
the stirring forces, by
\begin{align}
 I(r) 
 \label{eq:I_expr}
      &= 3 \int_0^\infty dk\ W(k)\ \left[ \frac{\sin{kr} - 
      kr\cos{kr}}{(kr)^3} \right] \ .
\end{align}
Here $I(r)$ is interpreted as the total energy injected into all scales $> 
r$.
Note that we may make the connection between $W(k)$ and the injection rate 
for the numerical simulations by
\beq
 \label{eq:I0}
 I(0) = \int^\infty_0 dk\, W(k) = \vep_W \ ,
\eeq
where the energy injection rate $\vep_W$ is defined in \eqref{forcing}.

If we were to apply \eqref{gen-khe} to freely-decaying turbulence, we
would set the input term
$I(r)$ equal to zero, to give:
\begin{equation}
 -\frac{3}{2}\frac{\partial U^2}{\partial t} = -\frac{3}{4}\frac{\partial 
 S_2}{\partial t} -\frac{1}{4r^4} \frac{\partial}{\partial r} \Big( r^4 S_3 
 \Big) + \frac{3\nu_0}{2r^4} \frac{\partial}{\partial r}\left( r^4 
 \frac{\partial S_2}{\partial r} \right) \ .
\end{equation}
Of course, for the case of free decay, we may also set $(3/2)\partial 
U^2/\partial t =-\vep$, after which we obtain the form of the KHE which is
familiar in the literature (e.g. see \cite{Monin75}). However, this can lead 
to problems if this substitution is retained for forced turbulence, for 
which it is not valid.

If, on the other
hand, we are considering forced turbulence which has reached a
stationary state, then we may set $\partial U^2/\partial t = \partial
S_2/\partial t =0$, whereupon (\ref{gen-khe}) reduces to the appropriate
KHE for forced turbulence,
\begin{equation}
 \label{eq:fKHE}
 I(r) =  -\frac{1}{4r^4} \frac{\partial}{\partial r} \Big( r^4 S_3(r) \Big) 
 + \frac{3\nu_0}{2r^4} \frac{\partial}{\partial r}\left( r^4 \frac{\partial 
 S_2(r)}{\partial r} \right) \ .
\end{equation}

As an aside, we note that this form for the forced KHE has several important 
differences from other approaches which have appeared in the literature 
\cite{Sirovich94,Gotoh02}. Previous approaches incorrectly retained the 
dissipation rate in the equation and essentially introduced an approximate
\emph{ad hoc} `correction' in order to take account of the 
forcing. This is, for example, presented for the third-order structure function as
\begin{equation}
 S_3(r) = -\frac{4\varepsilon r}{5} + Z(r) + 6\nu_0\frac{\partial 
 S_2}{\partial r} 
 \ ,
\label{zeqn}
\end{equation}
where $Z(r)$ is the \emph{ad hoc} correction \cite{Gotoh02}.
In contrast, we note that the origin of $\varepsilon$ in the KHE was 
$\partial U^2/\partial t$, which is zero for a stationary system, and 
instead show how its role is now played by the energy input function, 
$I(r)$. 
Thus, in our approach, instead of equation (\ref{zeqn}), we have
\begin{equation}
 S_3(r) = -\frac{4}{r^4} \int_0^r dy\ y^4 I(y) + 6\nu_0\frac{\partial 
 S_2}{\partial r} \ ,
\end{equation}
where $I(r)$ is calculated directly from the work spectrum, and is not 
approximated. Taking the limit $r\to0$ in equation \eqref{eq:I_expr}, for 
small scales we measure $I(r) = \varepsilon_W = \varepsilon$, and so
recover the Kolmogorov form of the KHE equation \cite{Kolmogorov41b}.

Returning to our form of the forced KHE, equation \eqref{eq:fKHE}, we now introduce the 
dimensionless structure functions $h_n(\rho)$
which are given by
\begin{equation}
 S_n(r) = U^n h_n(\rho) \ ,
\end{equation}
where $\rho = r/L$. Substitution into \eqref{eq:fKHE} leads to
\begin{align}
 -\frac{1}{4\rho^4} \frac{\partial}{\partial \rho} \Big(
\rho^4 h_3(\rho) \Big) \frac{U^3}{L} \equiv A_3(\rho|\RL) \frac{U^3}{L} \ ; 
\\
 \frac{3}{2\rho^4} \frac{\partial}{\partial
\rho}\left( \rho^4 \frac{\partial h_2(\rho)}{\partial \rho} \right)
\frac{\nu_0 U^2}{L^2} \equiv \frac{A_2(\rho|\RL)}{\RL} \frac{U^3}{L} \ ,
\end{align}
with $R_L = UL/\nu_0$ the Reynolds number based on the integral scale.
This introduces the coefficients $A_3$ and $A_2$, which are readily seen to 
be
\begin{align}
 A_3(\rho|\RL) &= -\frac{1}{4\rho^4} \frac{\partial}{\partial \rho} \Big(
\rho^4 h_3(\rho) \Big) \nonumber \\
 \label{eq:A2A3}
 A_2(\rho|\RL) &= \frac{3}{2\rho^4} \frac{\partial}{\partial
 \rho}\left( \rho^4 \frac{\partial h_2(\rho)}{\partial \rho} \right) \ .
\end{align}
Then, with some
rearrangement, the forced KHE \eqref{eq:fKHE} takes the dimensionless form
\begin{equation}
 I(\rho) \frac{L}{U^3} = A_3(\rho|R_L) + \frac{A_2(\rho|R_L)}{R_L} \ .
\label{dim-input}
\end{equation}
This simple scaling analysis has extracted the integral scale as the 
relevant lengthscale, and $R_L$ as the appropriate Reynolds number, for 
studying the behaviour of $\Ceps$. This was noted by Batchelor 
\cite{Batchelor53}, despite which it has become common practice to study 
$\Ceps = \Ceps(R_\lambda)$, as demonstrated by Fig. \ref{fig:Ceps_Rl}.

The input term may be expressed as an amplitude and a dimensionless shape 
function,
\begin{equation}
 \label{eq:phi}
 I(\rho) = \varepsilon_W \phi(\rho) \ ,
\end{equation}
where $\phi(\rho)$ contains all of the scale-dependent information and, as 
required by equation \eqref{eq:I0},
satisfies $\phi(0) = 1$.

\subsection{The limit of $\delta(\vec{k})$-forcing}

\begin{figure}
 \begin{center}
  \includegraphics[width=0.45\textwidth]{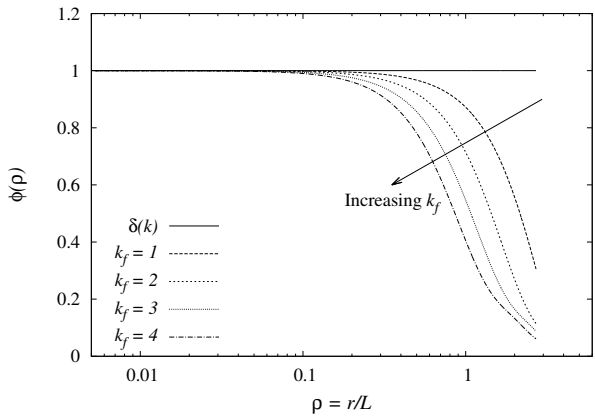}
 \end{center}
 \caption{The dimensionless input shape function $\phi(\rho)$, as defined by 
 equation \eqref{eq:phi}. The effect of varying the forcing band, $0 < k < 
 k_f$, is illustrated, showing the limit of $\delta(\vec{k})$-forcing.  
 Presented for $R_\lambda = 276$ data.}
 \label{fig:forcing}
\end{figure}

Figure \ref{fig:forcing} illustrates the shape of $\phi(\rho)$ and shows the 
effect of varying the forcing band defined in equation \eqref{forcing}, 
using data from our $R_\lambda = 276$ run.
As we reduce the width of the forcing band, we approach the limit of
$\delta$-function forcing in wavenumber space, corresponding to $\phi(\rho) 
= 1\ \forall\rho$. This cannot be studied using DNS, since the zero mode is 
not coupled to any other mode (and indeed is symmetry-breaking). But, for 
theoretical convenience, we consider the limit
analytically (or, alternatively, restrict our attention to scales for which 
$\phi(\rho) \simeq 1$) before addressing the complication added by scale 
dependence.

Now let us consider the \emph{dimensionless} KHE for the case of
$\delta(\mathbf{k})$-forcing, where $I(\rho) =  \vep_W = \vep$. Equation 
\eqref{dim-input} becomes
\begin{equation}
 \frac{\vep_W L}{U^3} = A_3(\rho|R_L) + \frac{A_2(\rho|R_L)}{R_L} \ ,
\end{equation}
from which, since $\vep = \vep_W$ and using equation \eqref{dim-diss}, we 
have
\begin{equation}
 \Ceps = A_3(\rho|R_L) + \frac{A_2(\rho|R_L)}{R_L} \ .
 \label{pre-DA-model}
\end{equation}
From the well known phenomenology associated with Kolmogorov's
inertial-range theories \cite{Kolmogorov41b}, as the Reynolds number
tends to infinity, we know that we must have $A_2/\RL \to 0$ and $A_3
\to \Cinf \equiv \mbox{constant}$. Hence, this equation suggests the
possibility of a simple model of the form
\begin{equation}
 \label{eq:DA_model}
 \Ceps = \Cinf + \frac{C_L}{R_L} \ ,
\end{equation}
where $\Cinf$ and $C_L$ are constants. 

Equation \eqref{pre-DA-model} can also be rewritten as
\begin{equation}
 \varepsilon = A_3(\rho|R_L)\frac{U^3}{L} + 
 A_2(\rho|R_L)\frac{\nu_0U^2}{L^2} \ .
\end{equation}
The first term on the RHS is essentially the Taylor surrogate, while the 
second term is a viscous correction. It has been shown \cite{McComb10b} 
that, for the case of decaying turbulence, the surrogate $U^3/L$ behaves 
more like a lumped-parameter representation for the maximum inertial 
transfer, $\varepsilon_T$, than the dissipation rate. The same is shown 
later for forced turbulence in Fig.  \ref{fig:Taylor_surrogate}, since the 
input rate (hence $\varepsilon$) is kept constant. Thus, the forced KHE is 
expressing the equivalence of the rate at which energy is transferred and 
dissipated (or injected) as $\nu_0 \to 0$. At finite viscosity, there is a 
contribution to the dissipation rate which has not passed through the 
cascade. In terms of our model equation,
\begin{equation}
 \varepsilon = \Cinf\frac{U^3}{L} + \nu_0 C_L\frac{U^2}{L^2} \to 
 \varepsilon_T \qquad\text{as}\qquad \nu_0 \to 0 \ ,
\end{equation}
where, from equation (\ref{eq:A2A3}), the asymptotic value is given by the expression
\begin{equation}
 \Cinf = \lim_{\nu_0 \to 0} A_3(\rho\vert R_L)  = -\frac{L}{U^3}\lim_{\nu_0 \to
0} \frac{1}{4r^4}\frac{\partial}{\partial r}\left(r^4 S_3(r) \right)   \ .
\end{equation}

\subsection{Modelling the scale dependence of coefficients with an \emph{ad 
hoc} profile function}

We now address the fact that the coefficients  $A_3(\rho)$ and $A_2(\rho)$ 
are \emph{not}
constants. They are separately scale-dependent; and, in general, may
also have a parametric dependence on the Reynolds number.

To begin, we use $\vep_W = \vep$ to rewrite equation \eqref{eq:phi} as 
$I(\rho)L/U^3 = \Ceps \phi(\rho)$, such that equation \eqref{dim-input} 
becomes
\begin{equation}
 \Ceps = \frac{A_3(\rho|R_L)}{\phi(\rho)} + \frac{A_2(\rho|R_L)}{R_L 
 \phi(\rho)} \ .
 \label{pre-DA-smodel}
\end{equation}
However, the fact that the left hand side of \eqref{pre-DA-smodel} is 
constant with respect to the dimensionless scale $\rho = r/L$ means that the 
separate dependences on $\rho$ on the right hand side must cancel.
In order to separate out the scale-dependent effects, we seek
semi-empirical decompositions for $A_3(\rho)$ and $A_2(\rho)$ which
satisfy the following conditions:
\begin{enumerate}
 \item The $\rho$-dependence of the terms on the RHS of 
 (\ref{pre-DA-smodel}) must cancel, since the LHS is a constant;
 \item As $\rho \to 0$, we have $A_3(\rho)/\phi(\rho) \to 0$ and so 
 $A_2(\rho)/R_L\phi(\rho) \to \Ceps(R_L)$ (it is entirely viscous);
 \item As $\rho \to \infty$, we have $A_2(\rho) \to 0$ and 
 $A_3(\rho)/\phi(\rho) \to \Ceps(R_L)$ (it is entirely inertial);
 \item As $R_L \to \infty$: $\Ceps(R_L) \to \Cinf = \text{constant}$.
\end{enumerate}

It is easily verified that these constraints are satisfied by the following
expressions, \begin{align}
  \label{eq:fitA3}
 \frac{A_3(\rho|R_L)}{\phi(\rho)} &= \Ceps \left[ 1 - H(\rho)
 \right]; \quad \text{and} \\
 \label{eq:fitA2}
 \frac{A_2(\rho|R_L)}{R_L\phi(\rho)} &= \Ceps H(\rho) \ ,
\end{align}
where we have introduced an \emph{ad hoc} profile function $H(\rho)$,
which in general must satisfy the conditions:
\beq
\lim_{\rho\to 0} H(\rho) =1 \quad \mbox{and} \quad \lim_{\rho \to
\infty} H(\rho) =0.
\eeq

The behaviour of the profile function at small and intermediate scales is 
also constrained by our knowledge of the structure functions. At small 
scales, the structure functions behave as $S_n \sim r^n$, which implies that 
$H(\rho) \simeq 1 - a\rho^2$ for some $a$. For large enough Reynolds 
numbers, in the inertial range of scales $S_2 \sim r^\gamma$ which leads to 
$H(\rho) \sim \rho^{\gamma-2}$, with $\gamma(R_L) \to 2/3$ as $R_L$ is 
increased.
Based on these additional constraints, we have chosen a suitable profile 
function to represent the scale dependence to be
\begin{equation}
H(\rho)= \left[ 1 + \frac{a\rho^2}{1+b\rho^\gamma} \right]^{-1}\ ,
\label{def-H}
\end{equation}
where $a$, $b$ and $\gamma$ are Reynolds number dependent and obtained by 
fitting to numerical results. We note that the actual values of these fit 
parameters do not affect our model \eqref{eq:DA_model} since the scale 
dependence cancels out.

\section{Numerical results}

\begin{figure}
 \begin{center}
  \includegraphics[width=0.45\textwidth]{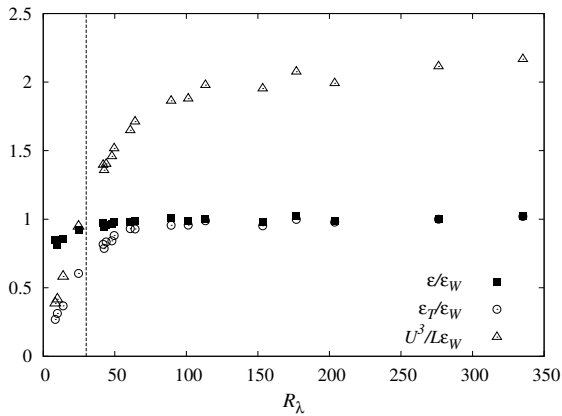}
 \end{center}
 \caption{ Variation with Taylor-Reynolds number of the
dissipation rate $\vep$, maximum inertial transfer $\vep_T$ and Taylor
surrogate $U^3/L$. Values to the left of the dashed line should be treated 
with caution: see the discussion in Section IV regarding Fig. \ref{fig:lowR}.}
 \label{fig:Taylor_surrogate}
\end{figure}
In Fig. \ref{fig:Taylor_surrogate} we show separately the behaviour of
the dissipation rate $\vep$, the maximum inertial flux $\vep_T$ and the
Taylor surrogate $U^3/L$, where each of these quantities was scaled on
the constant injection rate $\vep_W$. This is the basis of our first
observation. We see that the decrease of $\Ceps$, with increasing
Reynolds number, is caused by the increasing value of the surrogate in
the denominator, rather than by decay of the dissipation rate in the
numerator, as this remains fixed at $\vep = \vep_W$. This is the exact
opposite of the case for freely decaying turbulence, where the actual
dissipation rate decreases with increasing Reynolds number, while the
surrogate remains fairly constant \cite{McComb10b}. The figure also
shows how $U^3/L$ is a better lumped-parameter representation for 
$\varepsilon_T$ than $\varepsilon$ and that $\varepsilon/\varepsilon_T \to 
1$ from above as the Reynolds
number is increased, corresponding to the onset of an inertial range
\cite{McComb90a}.

\begin{figure}
 \begin{center}
  \includegraphics[width=0.45\textwidth]{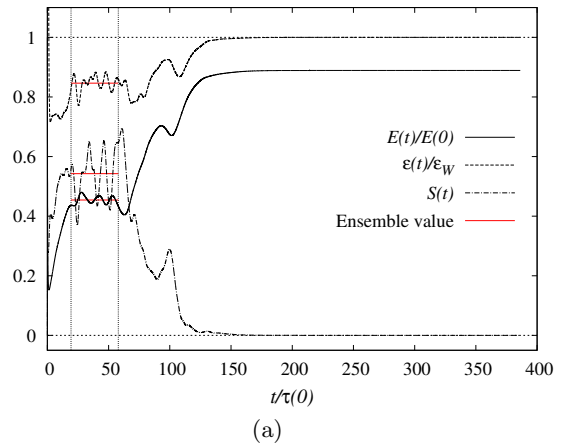} \\
  (a)\\
  \includegraphics[width=0.45\textwidth]{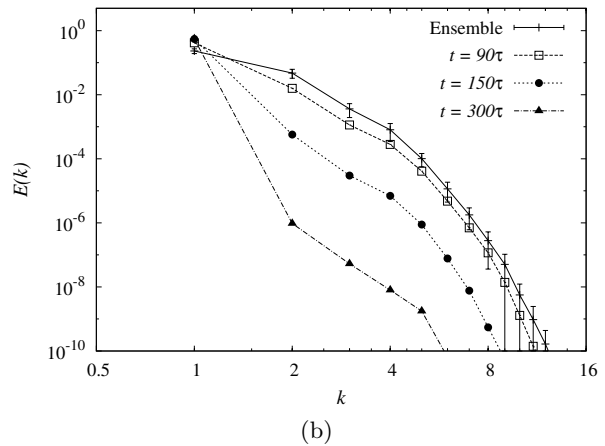} \\
  (b)
 \end{center}
 \caption{(a) Variation in time of the dissipation rate, total energy and 
 skewness, $S$, for our lowest value of the Reynolds number, $R_\lambda = 
 8.40$. The times usually associated with the steady state are indicated by 
 vertical dotted lines, with the ensemble values highlighted. For $t > 
 150\tau(0)$, we see a non-turbulent stable state.
 (b) Evolution of the energy spectrum from the ensemble averaged region 
 through the transition to a stable, non-turbulent state. Error bars 
 represent 3 standard deviations.}
 \label{fig:lowR}
\end{figure}
Note that all but the lowest two Reynolds number simulations conserved 
energy to within one standard deviation ($\sigma$) of the dissipation rate.  
However, runs with $R_\lambda < 25$ (indicated by the vertical dashed line 
in Fig.  \ref{fig:Taylor_surrogate}) should be treated with caution. A 
significant deviation from $\varepsilon = \varepsilon_W$ in a stationary 
simulation is an indication that the simulation is yet to reach steady 
state. A simulation to determine the long-time properties of these low 
Reynolds number runs was performed, with interesting results. As shown in 
Fig.  \ref{fig:lowR}(a), after the time usually associated with the steady 
state (indicated by vertical dotted lines), the simulation developed into a 
stationary stable state. This non-turbulent state has zero skewness, and 
essentially involves only one excited wavenumber, $k = 1$; see Fig.  
\ref{fig:lowR}(b). The ensemble averaged energy spectrum has been calculated 
within the times indicated by vertical dotted lines in Fig.  
\ref{fig:lowR}(a). Also plotted are the energy spectra at $t = 90\tau(0),\ 
150\tau(0)$ and $300\tau(0)$, corresponding to times within, towards the end 
of, and after the transition from pseudo-steady state to non-turbulent 
stable state, respectively. We see the development of a single-mode energy 
spectrum, with all the energy eventually being contained in the mode $k = 
1$.

This phenomenon has important consequences for the validity of all forced 
DNS results employing negative-damping, not just our own. It is currently 
unclear whether or not all Reynolds numbers will eventually develop into a 
stable, non-turbulent state, and one always measures a transient state 
masquerading as a steady state in which $\varepsilon$ fluctuates around a 
mean value which approaches $\varepsilon_W$ as Reynolds number is increased.

If instead this non-turbulent state is a low Reynolds number property, an 
alternative explanation for measuring $\varepsilon < \varepsilon_W$ involves 
the resolution of the large scales. It is becoming increasingly common to 
note that we do not only need to ensure that DNS is resolving the  small, 
dissipative scales, but also the large, energy containing scales, such as 
$L$. It is possible that this apparent lack of conservation of energy is 
caused by $L/L_\text{box}$ too large.

Further investigation is clearly needed. Until such information is 
available, we follow the literature and continue to use our DNS data for 
$R_\lambda > 25$. Despite simulations with lower Reynolds numbers being 
reported in the literature ($R_\lambda = 8$ \cite{Donzis05}) without energy 
conservation necessarily having been verified, our data corresponding to 
$R_\lambda < 25$ will not be taken into account on the basis that, for 
whatever reason, the simulation did not conserve energy.

\begin{figure}
 \begin{center}
  \includegraphics[width=0.45\textwidth]{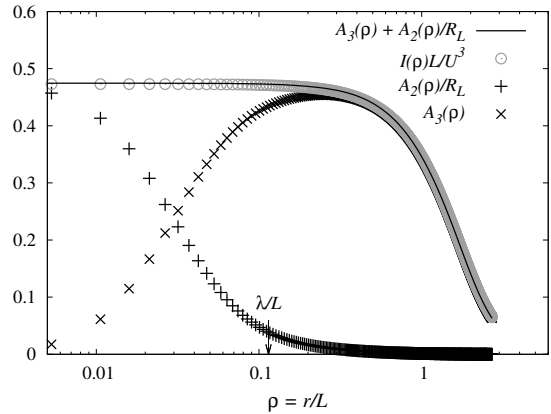}
 \end{center}
 \caption{ Dimensionless energy balance in the
Karman-Howarth equation, as expressed by equation \eqref{dim-input}.  
$R_\lambda = 276$. The Taylor microscale is labelled for comparison.}
 \label{fig:KHE_balance_dim}
\end{figure}

Figure \ref{fig:KHE_balance_dim} shows the balance of energy represented
by the dimensionless equation given as \eqref{dim-input}. For small scales 
($\rho < 0.2$ for the case $R_\lambda = 276$ shown) the input term satisfies 
$I(r) \simeq \varepsilon_W =
\varepsilon$, as expected since such scales are not directly influenced by 
the
forcing. We note that the second- and third-order structure functions may be 
obtained from the energy and transfer spectra, respectively, using
\begin{align}
\label{s2s3}
 S_2(r) &= 4 \int dk\ E(k)\ a(kr); \nonumber \\
 S_3(r) &= 12 \int dk\ \frac{T(k)}{k^2} \frac{\partial a(kr)}{\partial r} \ 
 ,
\end{align}
where the function $a(x)$ is:
\begin{equation}
 a(x) = \frac{1}{3} - \frac{\sin{x} - x\cos{x}}{x^3},
\end{equation}
with the derivatives of $a(kr)$ calculated analytically. This procedure
was introduced by Qian \cite{Qian97,Qian99} and more recently used by
Tchoufag \emph{et al} \cite{Tchoufag12}: the underlying transforms may
be found in the book by Monin and Yaglom \cite{Monin75}: equations
(12.75) and (12.141$'''$). From these expressions, the non-linear and
viscous terms, $A_3$ and $A_2$  given by equation \eqref{eq:A2A3}, have
been calculated using:
\begin{align}
 A_3(\rho) &= -\frac{3L}{U^3} \int_0^\infty dk\ T(k) \left[ 
 \frac{\sin{kL\rho} - kL\rho\cos{kL\rho}}{(kL\rho)^3} \right] \nonumber \\
 A_2(\rho) &= \frac{6\nu_0L}{U^3} \int_0^\infty dk\ k^2 E(k) \left[ 
 \frac{\sin{kL\rho} - kL\rho\cos{kL\rho}}{(kL\rho)^3} \right] .
\end{align}

\begin{figure}[!tb]
 \begin{center}
  \includegraphics[width=0.45\textwidth]{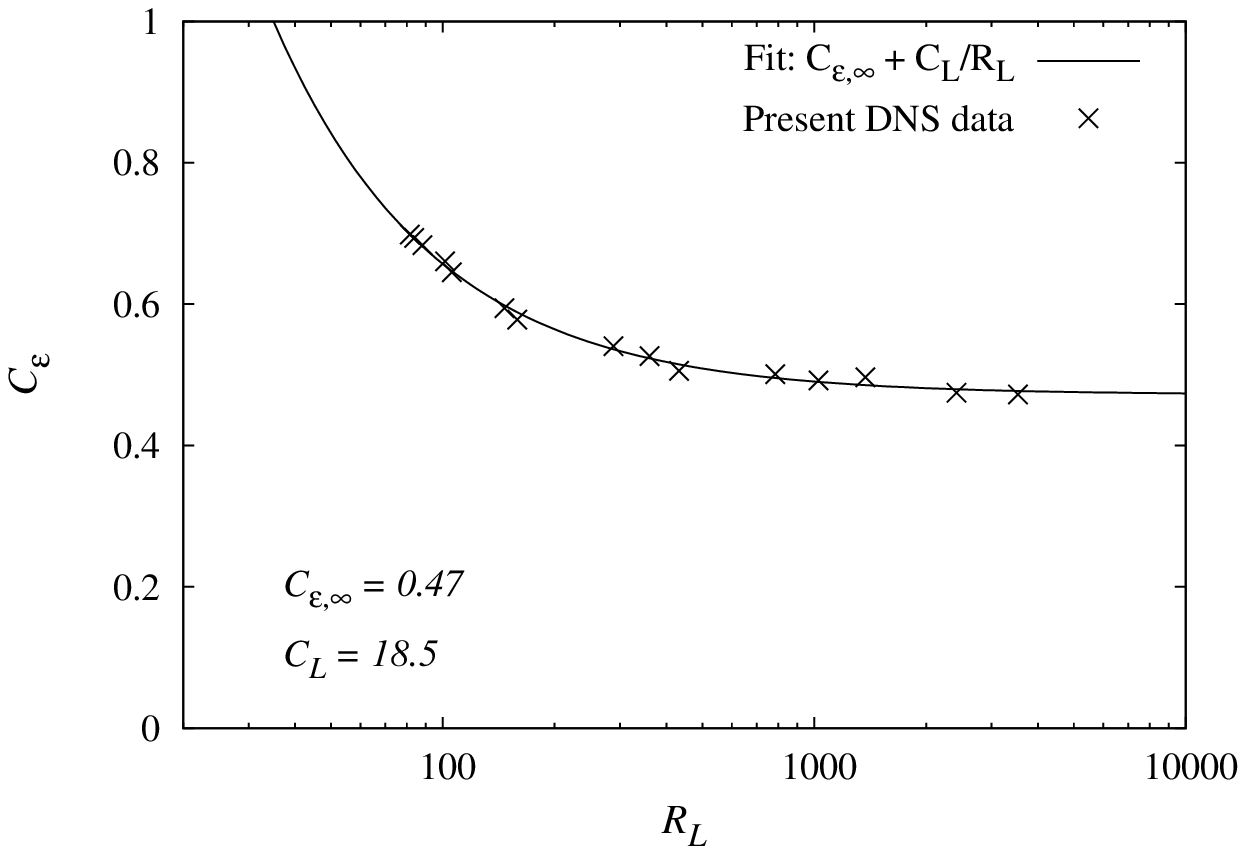}\\
  (a)\\
  \includegraphics[width=0.45\textwidth]{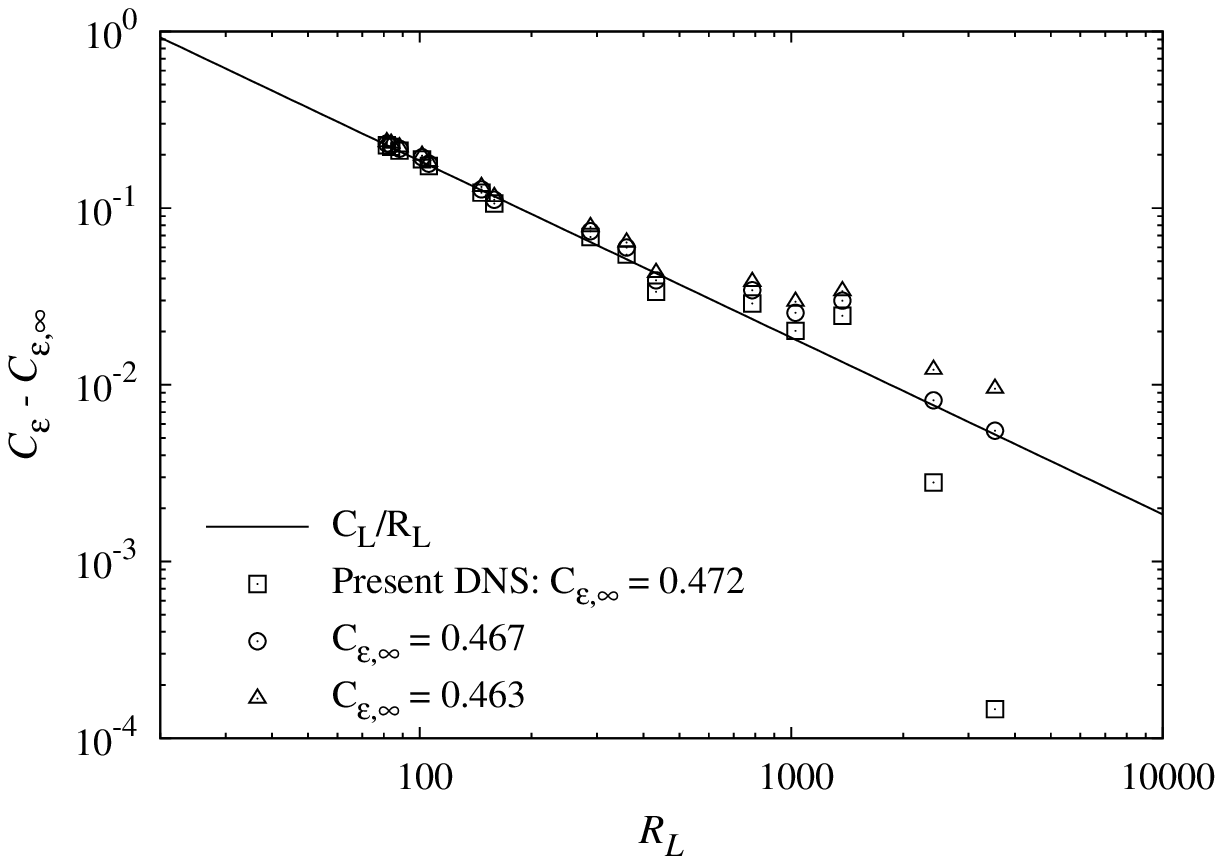}\\
  (b)
 \end{center}
 \caption{ (a) The expression given in equation \eqref{eq:DA_model} fitted 
 to present DNS data. (b) Log-log plot of the present DNS results for 
 $\Ceps$ against
Reynolds number, once the estimate of the
asymptote is subtracted. The solid line represents a slope of $-1.00$.
The effect of varying our estimate of the asymptote $\Cinf$ is shown by the 
three symbols.}
 \label{fig:DA_model}
\end{figure}

In order to test our model for the dimensionless dissipation rate, we fitted 
an expression of the form
(\ref{eq:DA_model}), but with an arbitrary power-law dependence $R_L^p$, to 
data
obtained with the present DNS, and it was found to agree very well, as shown 
in figure \ref{fig:DA_model}(a). The
exponent was found to be $p = -1.00 \pm 0.02$ and so supports the
model equation, with the constants given by $\Cinf = 0.47 \pm 0.01$ and $C_L 
=
18.5 \pm 1.3$.

A more graphic demonstration of this fact is given in Fig.  
\ref{fig:DA_model}(b).  The standard procedure of using a log-log plot to 
identify power-law behaviour is unavailable in this case, due to the asymptotic 
constant. For this reason, we subtracted the
estimated asymptotic value, and plotted $\Ceps - \Cinf$ against $\RL$ on
logarithmic scales. This allowed us to identify power-law behaviour 
consistent with $R_L^{-1}$. We also tested the effect of
varying our estimate of the value of the asymptote $\Cinf$. It can be
seen that the results were insensitive to this at the lower Reynolds 
numbers, where the $R_L^{-1}$ is being tested. At higher $\RL$, the viscous 
contribution represented by $C_L/R_L$ becomes negligible and instead we 
become strongly dependent on the actual value of $\Cinf$.

This model should be compared to other work in the literature. Sreenivasan 
\cite{Sreenivasan84} compared experimental decaying results to the 
expression for very low Reynolds numbers,
\begin{equation}
 \Ceps = \frac{15}{R_\lambda} \sqrt{\frac{\pi}{2}} \ .
\end{equation}
This used the isotropic relation $\varepsilon = 15\nu_0 U^2/\lambda^2$ 
(where $\lambda$ is the Taylor microscale) and the approximation $L/\lambda 
\simeq (\pi/2)^{1/2}$ \cite{Batchelor53}. Note that, while $15\sqrt{\pi/2} = 
18.8$, compared to $C_L = 18.5$ found in the present analysis, this 
expression involves $R_\lambda$ rather than $R_L$.

Later, Lohse \cite{Lohse94} used `variable range mean-field theory' to find 
an expression for the dimensionless dissipation coefficient by matching 
small $r$ and inertial range forms for the second-order structure function, 
and obtained
\begin{align}
 \Ceps 
  &= \Cinf \sqrt{1 + \frac{5b^3}{4R_\lambda^2}} \ ,
\end{align}
where $b = S_2(r)/(\vep r)^{2/3}$ such that $\Cinf = (h_2(1)/b)^{3/2}$. At 
low Reynolds numbers, the author reported $\Ceps = 18/R_L$. The asymptotic 
value was calculated by Pearson, Krogstad and van der Water 
\cite{Pearson02}, who used $h_2(1) \simeq 1.25$ and $b \simeq 2.05$,
to be $\Cinf \simeq 0.48$.

In an alternative approach, Doering and Foias \cite{Doering02} used the 
longest lengthscale affected by forcing, $l$, to derive upper and lower 
bounds on $\Ceps$,
\begin{align}
 \frac{4\pi^2}{\alpha^2 Re} \leq \Ceps \leq \left( \frac{a}{Re} + b \right) 
 \end{align}
for constants $a,b$, where $Re = Ul/\nu_0$ and $\alpha = L_\text{box}/l$.
While the upper bound resembles the present model, it is important to note 
that where these authors have obtained an inequality we have an equality.  
Based on Doering and Foias, an $R_\lambda$ form for the upper bound $A \big( 
1 + \sqrt{1 + (B/R_\lambda)^2} \big)$ was fitted to data by Donzis, 
Sreenivasan and Yeung \cite{Donzis05}, with $A \simeq 0.2$ and $B \simeq 92$ 
giving reasonable agreement, such that $\Cinf \simeq 0.4$.

Later still, Bos, Shao and Bertoglio \cite{Bos07} employed the idea of a
 finite cascade time to relate the expressions for $\Ceps$ in forced and
 decaying turbulence. Using a model spectrum, they then derived a form
for  $\Ceps$ and found the asymptotic value $\Cinf = 0.53$ with the
Kolmogorov  constant $C_K = 1.5$. Note that when we used their formula,
with the value $C_K = 1.625$ instead, this led to $\Cinf = 0.47$, as
found in the present work. With a simplified model spectrum, the 
authors then showed how their expression reduced to $\Ceps = 19/R_L$ for
low  Reynolds numbers (when $E(k) \sim k^4$ at low $k$) in agreement
with $C_L =  18.5$ found here (within one standard error).

\begin{figure}[!tb]
 \begin{center}
   \includegraphics[width=0.45\textwidth]{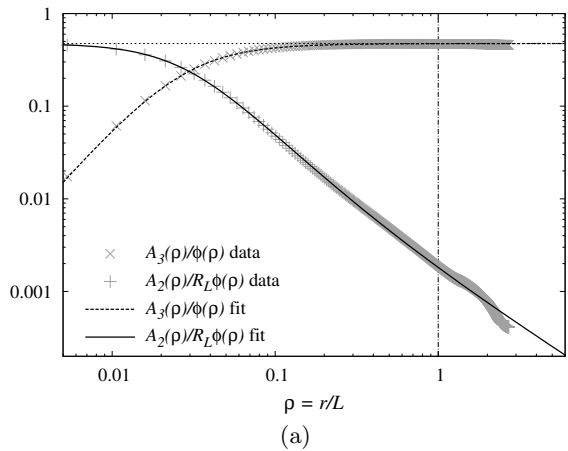} \\
   (a)\\
   \includegraphics[width=0.45\textwidth]{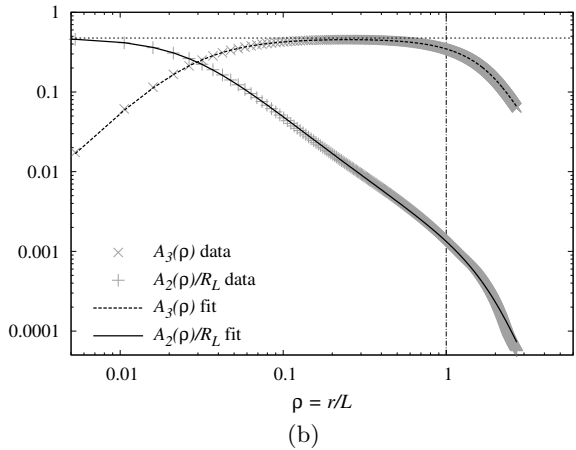} \\
   (b)
 \end{center}
 \caption{ (a) The fit for $A_2(\rho)/R_L\phi(\rho)$, as given by equation 
 \eqref{eq:fitA2}, to present DNS data. This also determines 
 $A_3(\rho)/\phi(\rho)$; see equation \eqref{eq:fitA3}.  The fit was 
 performed for the region $\rho \leq 1$. (b) The equivalent fit for 
 $A_3(\rho)$ and $A_2(\rho)/R_L$.  In both parts, $\Ceps$ is indicated by 
 the horizontal dotted line.}
 \label{fig:model_fit}
\end{figure}

The expression for $A_2(\rho)/R_L\phi(\rho)$ given by equation 
\eqref{eq:fitA2} was
fitted to the present DNS data to find $a,\ b$ and $\gamma$. This also fixed 
the
form for $A_3(\rho)/\phi(\rho)$, as given by equation \eqref{eq:fitA3}. The 
fit was
performed up to the integral scale, $\rho = 1$, as shown in Fig.  
\ref{fig:model_fit}(a) by the vertical dash-dot line, above which the 
simulations become less well resolved.
Clearly, agreement is excellent for $\rho < 1$. Figure 
\ref{fig:model_fit}(b) then uses the measured function $\phi(\rho)$ to plot 
the equivalent fit to DNS data for $A_3$ and $A_2/R_L$. The scale dependence 
of $A_2$ and $A_3$ is, therefore, well modelled by our choice of profile 
function, $H(\rho)$. As a consequence, the scale dependence in equation  
\eqref{pre-DA-smodel} cancels out in such a way that $\Ceps(R_L)$ can still 
be modelled using equation \eqref{eq:DA_model}, despite finite forcing 
introducing scale dependence $\phi(\rho)$ to the input term. One could 
therefore replace $\Ceps$ in equations \eqref{eq:fitA3} and \eqref{eq:fitA2} 
with $\big[\Cinf+C_L/R_L\big]$.

\section{Conclusions}

We have presented a new form of the KHE for forced turbulence which
differs  from that commonly found in the literature. In deriving this
equation from  the Lin equation, we have obtained a scale-dependent
energy input term  \eqref{eq:I_expr}.  Our new form of the general KHE,
equation  \eqref{gen-khe}, correctly reduces to the well-known form for
decaying  turbulence.

By scaling the forced KHE into a dimensionless form \eqref{dim-input}, we 
see that the appropriate Reynolds number for studying the variation of the 
dimensionless dissipation, $\Ceps$, is that corresponding to the integral 
scale, $R_L$. In the limit of $\delta(\vec{k})$-forcing, or for scales well 
below the influence of any forcing, the dimensionless equation suggests the 
simple model \eqref{eq:DA_model} for the balance of inertial and viscous 
contributions to the dimensionless dissipation rate. The new model has been 
fitted to the present DNS data with excellent agreement. It also shows that
the behaviour of the dimensionless dissipation rate, as found
experimentally, is entirely in accord with the Kolmogorov (K41) picture
of turbulence and, in particular, with Kolmogorov's derivation of his
`$4/5$' law \cite{Kolmogorov41b}, the one universally accepted result in
turbulence.

The scale dependence of the inertial and viscous terms, $A_3$ and $A_2$, 
caused by finite forcing have been shown to compensate one another exactly 
\eqref{pre-DA-smodel}, and as such have been modelled by a single profile 
function $H(\rho)$. The scale independence of equation \eqref{pre-DA-smodel} 
can then be used to motivate the application of the model given by equation 
\eqref{eq:DA_model} to general, finite forcing.

The authors thank Matthew Salewski for reading the manuscript and making
a number of helpful comments. AB and SY were funded by the STFC.
We thank one of the referees for drawing our attention to the statistically 
significant lack of energy conservation in simulations with $R_\lambda < 25$ 
and also for pointing out the implications of the small-scale limit for the 
\emph{ad hoc} profile function.

\end{document}